\documentclass[useAMS,usenatbib,usegraphicx]{mn2e}
\usepackage{color}

\usepackage{epsfig,graphicx,latexsym,amsmath,amssymb}
\usepackage{natbib}
\usepackage{hyperref}
\usepackage{mathrsfs}
\usepackage{lastpage}
\usepackage{bm}
\usepackage{verbatim}
\usepackage[labelformat=empty]{subfig}

\def\simlt{\lower.5ex\hbox{$\; \buildrel < \over \sim \;$}}
\def\simgt{\lower.5ex\hbox{$\; \buildrel > \over \sim \;$}}

\bibpunct[,]{(}{)}{;}{a}{}{,}

\title[Suprathermal electron firehose constraints]
{Firehose constraints for the solar wind suprathermal electrons}

\author[M. Lazar, S.M. Shaaban, S. Poedts 
and \v{S}. \v{S}tver\'{a}k]{M. Lazar,$^{1,2}$\thanks{E-mail: mlazar@tp4.rub.de} 
S.M. Shaaban,$^{2,3}$
S. Poedts$^{2}$ 
and \v{S}. \v{S}tver\'{a}k$^{4,5}$
\\
$^1$ Institut f\"ur Theoretische Physik, Lehrstuhl
IV: Weltraum- und Astrophysik, Ruhr-Universit\"at Bochum, D-44780
Bochum, Germany\\  $^2$ Center for Plasma Astrophysics, KU Leuven, Celestijnenlaan 200B, 3001
Leuven, Belgium\\  $^3$ Theoretical Physics Research Group, Physics Department, Faculty of Science, Mansoura University, 35516, Egypt\\ 
$^4$Institute of Atmospheric Physics, Czech Academy of Sciences, Prague, Czech Republic\\
$^5$Astronomical Institute, Czech Academy of Sciences, Ondrejov, Czech Republic\\}
\begin{document}

\date{Accepted  MM DD. Received 2016 MM DD; in original form 2016}

\pagerange{\pageref{firstpage}--\pageref{lastpage}} \pubyear{2014}

\maketitle

\label{firstpage}

\date{\today}

\begin{abstract}

The indefinite increase of temperature predicted by the solar wind expansion in the direction parallel to the interplanetary 
magnetic field is already notorious for not being confirmed by the observations. In hot and dilute plasmas from space 
particle-particle collisions are not efficient in constraining large deviations from isotropy, but the resulting 
firehose instability provides in this case plausible limitations for the temperature anisotropy of the thermal (core) 
populations of both the electron and proton species. The present paper takes into discussion the suprathermal (halo) 
electrons, which are ubiquitous in the solar wind. Less dense but hotter than the core, suprathermals may be highly 
anisotropic and susceptible to the firehose instability. 
The main features of the instability are here derived from a first-order theory for conditions specific to the 
suprathermal electrons in the solar wind and terrestrial magnetospheres. Unveiled here, new regimes of the electron firehose
instability may be exclusively controlled by the suprathermals. The instability is found to be systematically 
stimulated by the suprathermal electrons, with thresholds that approach the limits of the temperature anisotropy 
reported by the observations. These results represent new and valuable evidences for the implication of the firehose instability
in the relaxation of the temperature anisotropy in space plasmas.

\end{abstract}

\begin{keywords}
Sun: solar wind --- electrons --- temperature anisotropy --- 
electromagnetic instabilities --- methods: analytical; observational
\end{keywords}

\section{Introduction}


In collision-poor plasmas from space large deviations from thermodynamic equilibrium cannot be relaxed by the 
particle-particle (Coulomb) collisions, but can presumably be constrained by the resulting kinetic instabilities.
Thus, if the solar wind expands adiabatically the CGL invariants conserve \citep{chew56} leading to an indefinite 
increase of temperature in the direction parallel to the inteplanetary magnetic field, i.e., $T_\parallel > T_\perp$. 
However, the in-situ measurements do not confirm such an increase of their parallel temperature with heliocentric 
distance, but indicate bounds of the temperature anisotropy of plasma particles \citep{Kasper02, Hellinger06, 
St2008}. Because collisions are not efficient, the most invoked mechanism that can limit the increase of parallel 
temperature is the firehose instability \citep{Eviatar70,Kasper02,Hellinger06, St2008, La2014}. 

The firehose instability driven by the anisotropic electrons with $A \equiv T_\perp / T_\parallel < 1$, also known as the 
electron firehose instability (EFHI), is particularly important as it can mediate a resonant transfer of (free) energy from 
electrons to protons \citep{Paesold99, Messmer02}. This energy transfer from small to large scales 
is facilitated by the quasi-parallel EFH modes, which are left-handed (LH) circularly polarized and have characteristic
frequencies and growth rates in the range of the proton cyclotron frequency. Besides the propagating 
(non-zero frequency) modes predominant at small angles (quasi-parallel) with respect to the magnetic field direction, 
the firehose instability may destabilize an additional aperiodic (non-propagating) branch which exists only for oblique 
directions \citep{Ga2003,Ca2008,Hellinger14}. Although it is well known that the suprathermal populations
are ubiquitous in the solar wind \citep{Lin1998,Pier2010,La2012a}, the anisotropic temperature is in general 
quantified by a bi-Maxwellian distribution function, which
is relevant only for the thermal core of the solar wind electrons. In this case the aperiodic FHI 
is found to grow faster than the propagating modes, and the instability thresholds approach well enough
the limits of the core anisotropy reported by the observations \citep{St2008}. For anisotropies exceeding these thresholds,
the free energy is dissipated by the resulting instabilities, which may also scatter particles back towards 
quasi-equilibrium states and prevent the anisotropy to grow \citep{Ga1994,Ga1998}. Instead, for the suprathermal
electrons from the solar wind the limits of their anisotropy are markedly departed from the instability 
thresholds derived for bi-Maxwellian populations, see Figure~6 and the analysis in \cite{St2008}. 
This disagreement may simply be motivated by the fact that
suprathermal populations cannot be properly described by the Maxwellian distribution functions, but they can be 
accurately reproduced by the Kappa power-laws \citep{Va1968,Ma2005,Pier2010}. Ubiquitous in the solar wind 
and subsequent planetary environments, e.g., terrestrial magnetosphere, see the review by \cite{Pier2010}, 
suprathermal electrons are more dilute but hotter than the core populations. The relaxation through the 
particle-particle collisions is even less efficient in this case, but kinetic instabilities are expected to explain 
the limits of temperature anisotropy reported by the observations.

In the present paper we propose a refined analysis of the suprathermal electrons by using a bi-Kappa distribution function to 
describe the anisotropy of these populations. In the limit of a high power-index $\kappa \to \infty$ 
the (bi-)Kappa distribution function reduces to a (bi-)Maxwellian. \cite{Ma2005} and \cite{St2008} have used the bi-Kappa
model to quantify the velocity distributions and the principal properties of the suprathermal electrons in the solar wind, 
e.g., the components of the anisotropic temperature, parallel ($T_\parallel$) and perpedicular ($T_\perp$) to 
the magnetic field direction. The suprathermal electrons are found to be highly anisotropic and with a predominant
excess of parallel temperature susceptible to the FHI. The bi-Kappa model was also extensively invoked in theories 
of dispersion and stability by adopting two alternative assumptions for the temperature of Kappa populations, to be either dependent or 
independent of the power-index $\kappa$. Studies of the FHI \citep{La2009,La2011} assume $\kappa$-independent 
temperatures, and find, contrary to the expectations, that the instability is inhibited by the suprathermals and 
the instability thresholds do not approach but depart even more from the anisotropy bounds of the solar wind suprathermal electrons.
However, from a recent analysis on the applicability of Kappa distributions \citep{La2015a,La2016} it 
becomes evident that a representation with a $\kappa$-dependent temperature may provide a more natural 
interpretation of the suprathermal populations for three fundamental reasons: (1)~it corresponds to a Maxwellian
limit which reproduces more accurately the thermal (core) population enabling for a direct and realistic
comparison \citep{La2015a}; (2)~the kinetic instabilities show a systematic stimulation in the presence of suprathermal 
electrons \citep{La2015a, Vinas2015,Sh2016} as one may expect from the excess of free energy acumulated by these populations;
and (3)~the observations show strong evidence of $\kappa$-dependent temperatures, which increase in 
the presence of suprathermal populations, i.e., temperatures increase with decreasing the power-index $\kappa$ \citep{Pier2016}.

%
%
Motivated by these premises, here we re-analyse the instability of the EFH mode by modeling the suprathermal 
electrons with a bi-Kappa approach with 
$\kappa$-dependent temperatures. In this preliminary analysis we restrict to the same parallel (non-zero frequency) 
modes studied before by \cite{La2009,La2011}. The bi-Kappa approach is introduced in section~2, enabling us
to derive the dispersion relation for the FHI modes. The main features of the instability, are derived
and discussed in section~3. In addition, the EFH thresholds are compared with the observations of the suprathermal 
electron anisotropy. The results of the present work are summarized in section 4.  

\section{Bi-Kappa electrons. Dispersion relations}

We first introduce the analytical model for the velocity 
distributions of suprathermal electrons detected in space 
plasmas \citep{Ma2005,St2008}. The suprathermal (halo) electrons are assumed to be 
a gyrotropic component (isotropic in the plane transverse to the magnetic field) 
with a bi-axis temperature anisotropy $T_\perp \ne T_\parallel$, where 
$\parallel$ and $\perp$ denote directions relative to the magnetic field. 
The distribution of suprathermal electrons in velocity space with polar coordinates 
$(v_{\perp} \cos \phi, v_{\perp} \sin \phi, v_{\parallel}) = (v_x, v_y, v_z)$ is described 
by a bi-Kappa distribution function
\begin{align}
F^\kappa (v_{\parallel}, v_{\perp}) = & {1 \over \pi^{3/2}
\theta_{\parallel} \theta_{\perp}^2} \, {\Gamma[\kappa] \over
\kappa^{1/2} \Gamma[\kappa - 1/2]}\notag \\
& \times \left(1 + {v_{\parallel}^2\over \kappa
\theta_{\parallel}^2 } + {v_{\perp}^2\over \kappa
\theta_{\perp}^2 }\right)^{-\kappa-1}, \label{e1}
\end{align}
which is normalized to unity, and where $\theta_{\parallel, \perp}$ 
are thermal velocities defined by, respectively, the parallel and perpendicular temperatures as moments of second order
\begin{align}
T^\kappa_{\parallel} = {m \over k_B} \int d{\bf v} v_\parallel^2
F^\kappa (v_{\parallel}, v_{\perp}) = {\kappa \over \kappa-3/2} 
{m \theta_{\parallel}^2\over 2 k_B}, \label{e2}
\end{align}
\begin{align}
T^\kappa_{\perp} ={m \over 2 k_B} \int d{\bf v} v_\perp^2
F^\kappa (v_{\parallel}, v_{\perp}) = {\kappa \over
\kappa-3/2} {m \theta_{\perp}^2 \over 2 k_B}. \label{e3}
\end{align}
The bi-Kappa simply reduces to a bi-Maxellian in the limit of a very large
$\kappa \to \infty$
\begin{align}
F^M (v_{\parallel}, v_{\perp}) = & {1 \over \pi^{3/2}
\theta_{\parallel} \theta_{\perp}^2} \, \exp \left(- {v_{\parallel}^2\over 
\theta_{\parallel}^2 } + {v_{\perp}^2\over \theta_{\perp}^2 }\right), \label{e4}
\end{align}
with
\begin{align}
T^M_{\parallel} = {m \over k_B} \int d{\bf v} v_\parallel^2
F^M(v_{\parallel}, v_{\perp}) = {m \theta_{\parallel}^2\over 2k_B} < T^\kappa_{\parallel} , \label{e5}
\end{align}
\begin{align}
T^M_{\perp} = {m \over 2 k_B} \int d{\bf v} v_\perp^2
F^M(v_{\parallel}, v_{\perp}) = {m  \theta_{\perp}^2\over 2 k_B} < T^\kappa_{\perp}. \label{e6}
\end{align}
Notice in case that the temperature of suprathermal electrons decreases with increasing the 
power-index $\kappa$ and reaches a minimum for the Maxwellian limit.

In the direction parallel to the magnetic field (${\bf k} \parallel
{\bf B}$), the electromagnetic (EM) modes are decoupled from the electrostatic
oscillations, and are described by the following general
dispersion relation \citep{Ga1993}
\begin{align}
{k^2c^2 \over \omega^2}= & 1  + {4 \pi \over \omega^2} \sum_a {e_a
\over m_a} \int_{-\infty}^{\infty} \, {dv_{\parallel} \over \omega -
k v_{\parallel} \pm \Omega_a}
\int_0^{\infty} \, dv_{\perp} \notag \\
& \times v_{\perp}^2 \left[(\omega - k v_{\parallel}) {\partial
F_{a} \over \partial v_{\perp}} + k v_{\perp} {\partial F_{a} \over
\partial v_{\parallel}} \right], \label{e7}
\end{align}
where $\omega$ and $k$ are respectively, the frequency and the
wavenumber of the plasma modes, $c$ is the speed of light in vacuum,
$\Omega_a = q_a B_0 / (m_a c)$ is the gyrofrequency for the particles of sort $a$, e.g., $a = e$ for electrons
and $a=p$ for protons, respectively, and "$\pm$" describes the 
circularly polarized EM modes with right-hand (RH) and 
left-hand (LH) polarizations, respectively. For the advanced model 
introduced in equation (\ref{e1}) the dispersion relation becomes
\begin{align}
{k^2c^2 \over \omega^2} = 1  & + \sum_a {\omega_{a,h}^2 \over \omega^2} 
\Big[A_a-1 \notag \\ & + {(A_a-1)(\omega \pm \Omega_a) + \omega \over k
\theta_{a,\parallel}} \, Z_{\kappa} \left(\omega \pm \Omega_a \over k
\theta_{a,\parallel} \right) \Big], \label{e8}
\end{align}
where $A_a = T_{a,\perp} /T_{a,\parallel}$ is the temperature
anisotropy, 
\begin{align}
Z_{\kappa}(f) = &{1 \over \pi^{1/2} \kappa^{1/2}} \, {\Gamma
(\kappa) \over \Gamma
\left( \kappa -{1 \over 2}\right)} \notag \\
& \times \int_{-\infty}^{\infty} dx \, {(1+x^2/\kappa)^{-\kappa }
\over x - f}, \;\;\; \Im(f)>0 \label{e9}
\end{align}
is the Kappa plasma dispersion function \citep{La2008} of
argument
\begin{align}
 f_{\kappa} = {\omega \pm \Omega_a \over k \theta_{a,\parallel}}.
\label{e10}.
\end{align}
In the Maxwellian limit this function reduces to the standard plasma
dispersion function \citep{Fr1961} 
\begin{align} 
Z(f) = & {1 \over \pi^{1/2}}
\int_{-\infty}^{+\infty} dx \, {\exp (-x^2) \over x - f}, \; \;\;
\Im (f) > 0 \label{e11} 
\end{align}
of argument
\begin{align}
f = {\omega \pm \Omega_a \over k w_{a}}. \label{e12}
\end{align}
Note that for our model introduced in Eqs.~(\ref{e1})--(\ref{e6}), the anisotropy does not depend on $\kappa$,
i.e., $A = T^\kappa_{\perp}/T^\kappa_{\parallel} = T^M_{\perp}/T^M_{\parallel}$. 

We investigate the EFHI, which is a LH EM mode driven unstable by an excess of electron temperature in parallel direction 
$T_{e,\parallel} > T_{e,\perp}$, i.e., $A_e < 1$. According to (\ref{e8}),
the dispersion relation describing these modes can be rewritten with normalized 
quantities as follows
\begin{eqnarray}
\mu \left[ A_{e}-1+ \frac{ A_e \left(\tilde{\omega}+\mu \right) -\mu }{\tilde{k}
\sqrt{\mu \beta_{e,\parallel}^M}} Z_\kappa\left( \frac{\tilde{\omega}+\mu }{\tilde{k}\sqrt{\mu \beta _{e,\parallel}^M}}\right) \right] \notag \\
+\frac{\tilde{\omega}}{\tilde{k} \sqrt{\beta _{e,\parallel}^M /\Theta}} Z\left( \frac{\tilde{\omega}-1}
{\tilde{k}\sqrt{\beta_{e,\parallel}^M / \Theta}}\right) =\tilde{k}^{2},  \label{e13}
\end{eqnarray}
where protons are assumed Maxwellian and isotropic $A_p = 1$, and $\tilde{\omega}=\omega /\Omega _{p}$,
$\tilde{k}=~kc/\omega_{p,p}$, $\mu =~m_{p}/m_{e}$ is the
proton/electron mass ratio, $\Theta=~$~$T_{e,\parallel}^M/T_{p,\parallel }^M\ $ is
the electron/ proton parallel temperature ratio in the Maxwellian
limit for both species, and $\beta_{e, \parallel}^M=~8\pi
n_{e}k_{B}T_{e,\parallel}^M/B_{0}^{2}$ is the parallel electron beta
parameter in the Maxwellian limit $\kappa \to \infty$. The dispersion relation for 
bi-Maxwellian distributed electrons can be obtained from Eq. (\ref{e13}) only by changing $Z_\kappa$
with the Maxwellian plasma dispersion function from (\ref{e11}).

\section{EFHI. Thresholds vs. suprathermal electron anisotropy}

We have solved the dispersion relation~(\ref{e13}) numerically, and analyzed the unstable firehose 
solutions. In this section we present the main features of the EFHI, namely, growth rates, wave-frequencies
and wave-numbers, as well as the anisotropy thresholds, and restrict our analysis only to the unstable 
regimes controlled mainly by the suprathermal electrons. The effects of suprathermal populations are triggered
by their temperature anisotropy and their abundance, which is quantified by the finite (especially low) values of the power-index $\kappa$.

\begin{figure}
\centering
    \includegraphics[width=80mm]{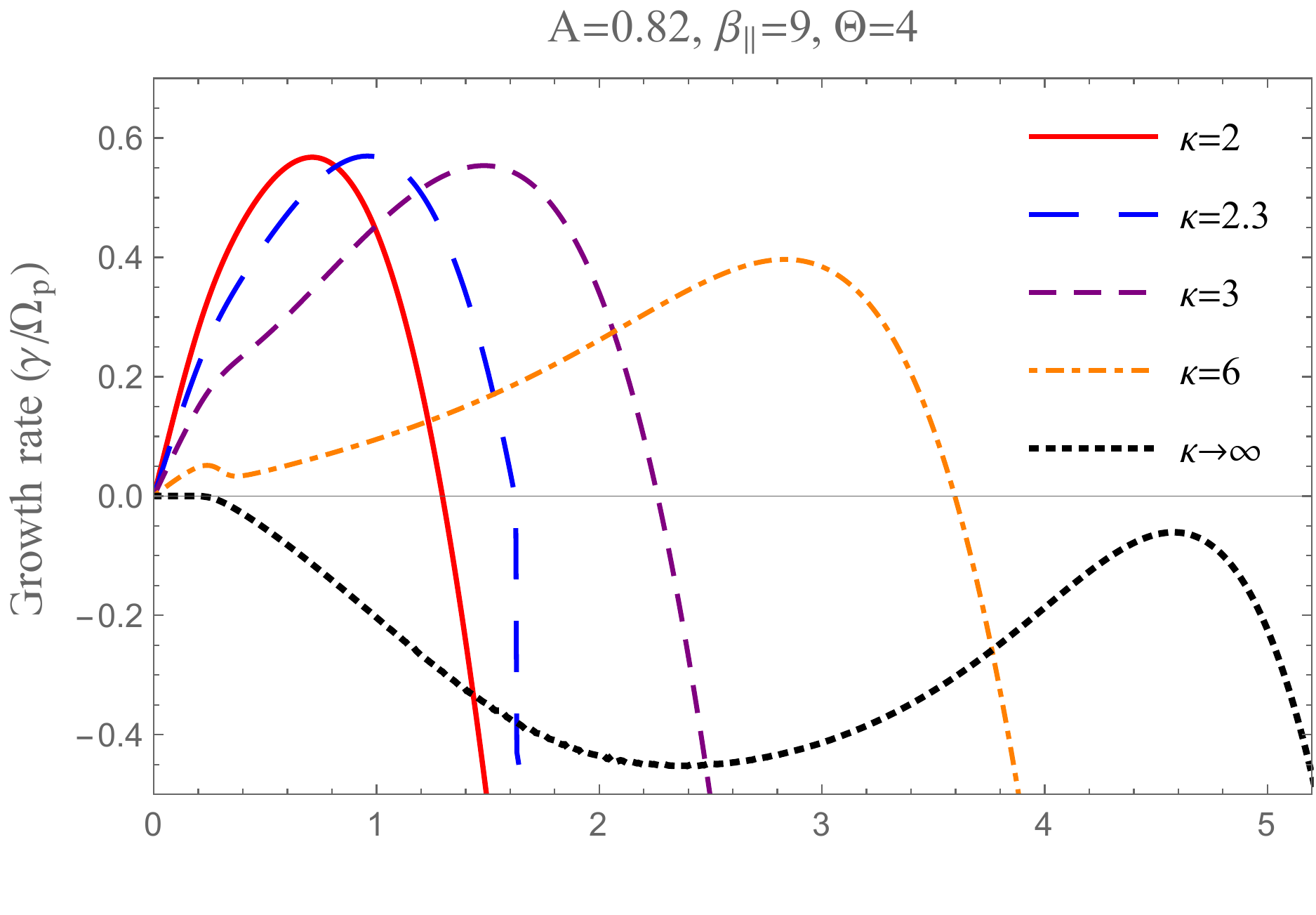}
    \includegraphics[width=80mm]{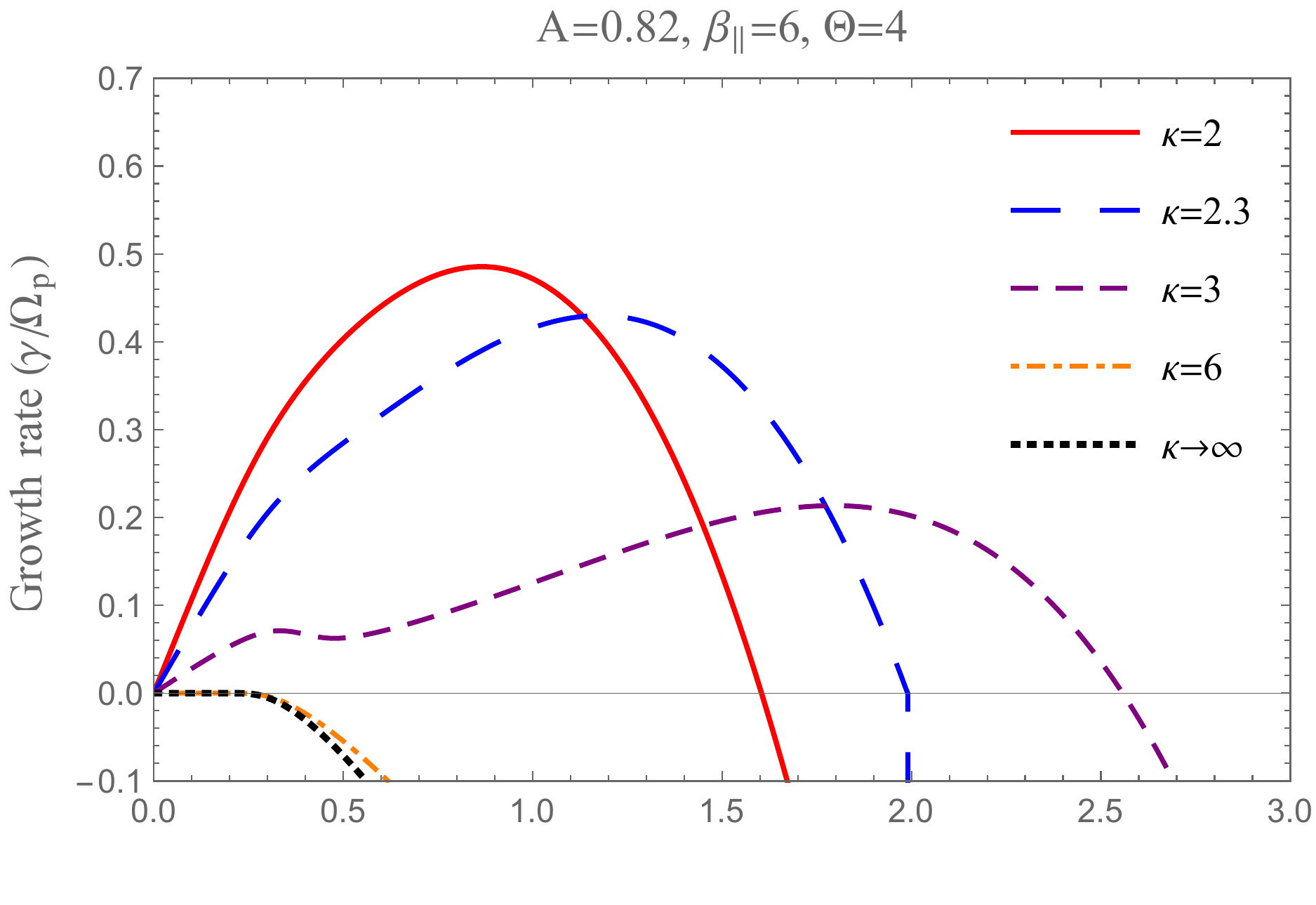}
    \includegraphics[width=80mm]{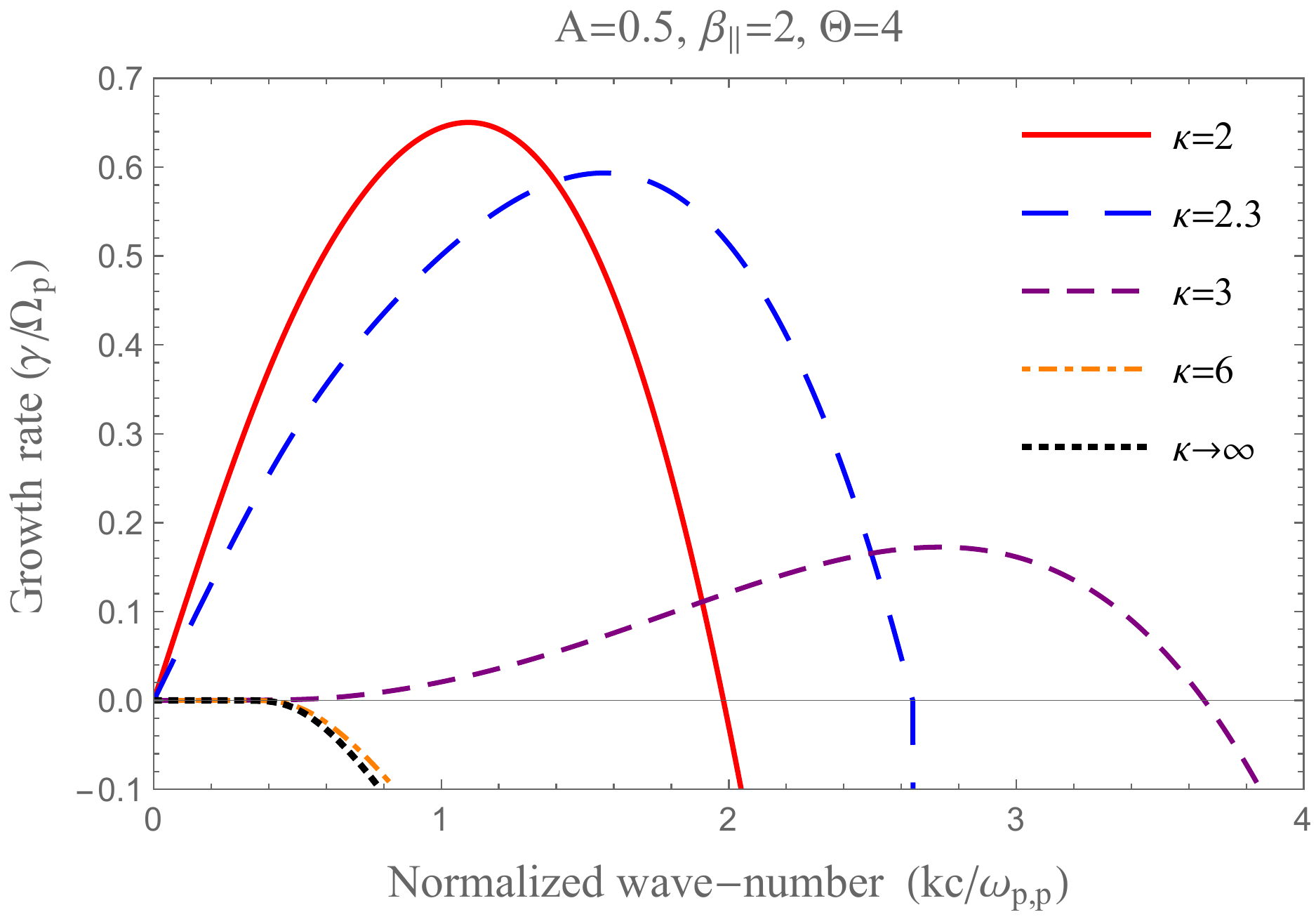}
    \caption{Effects of the suprathermal electrons quantified by the power-index $\kappa=$2, 2.3, 3, 6, 
    $\infty$, on the growth rates of the EFH instability for different plasma beta explicitly given in each
    panel.} \label{f1}%
\end{figure}

\begin{figure}
\centering
    \includegraphics[width=80mm]{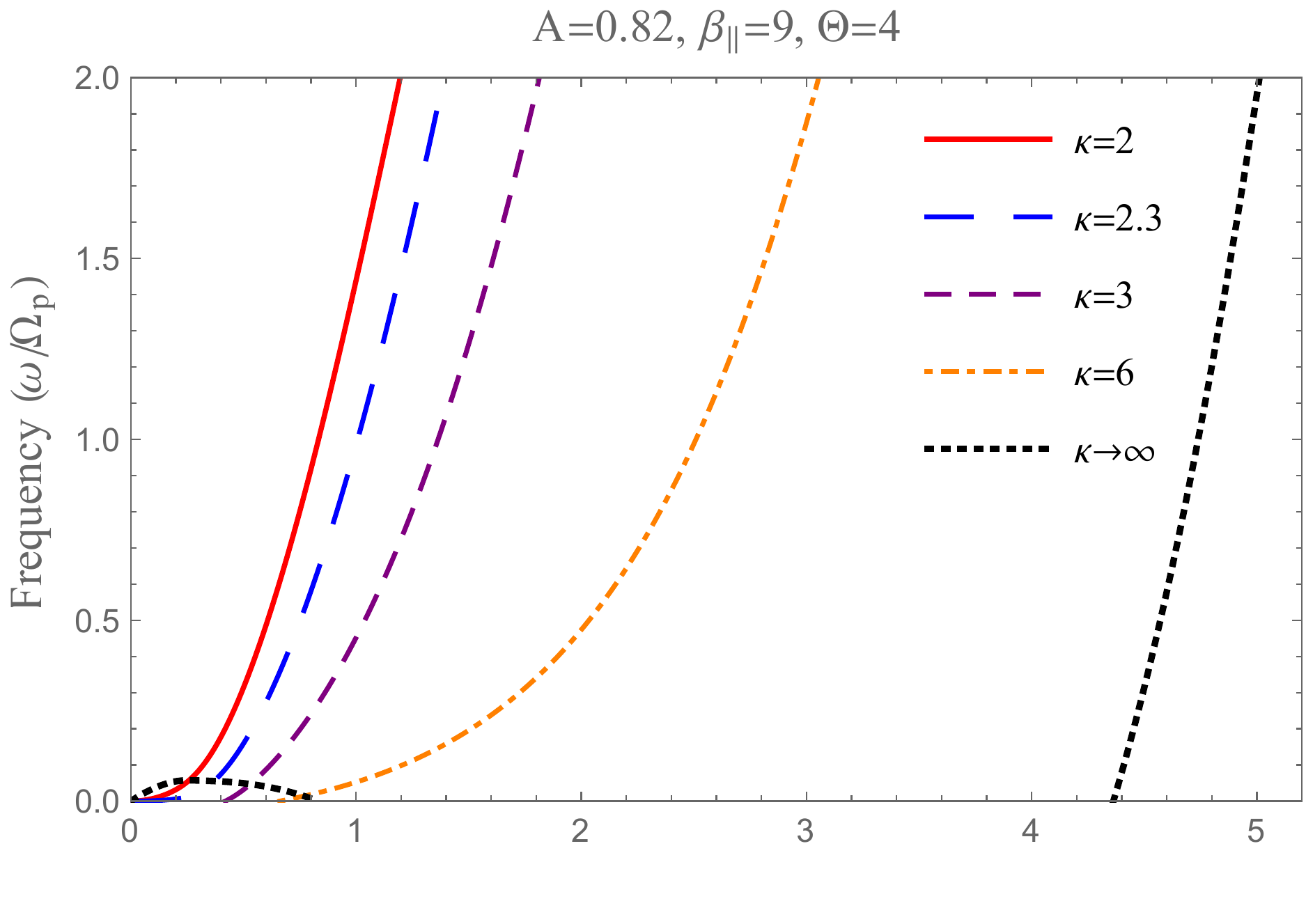}
    \includegraphics[width=80mm]{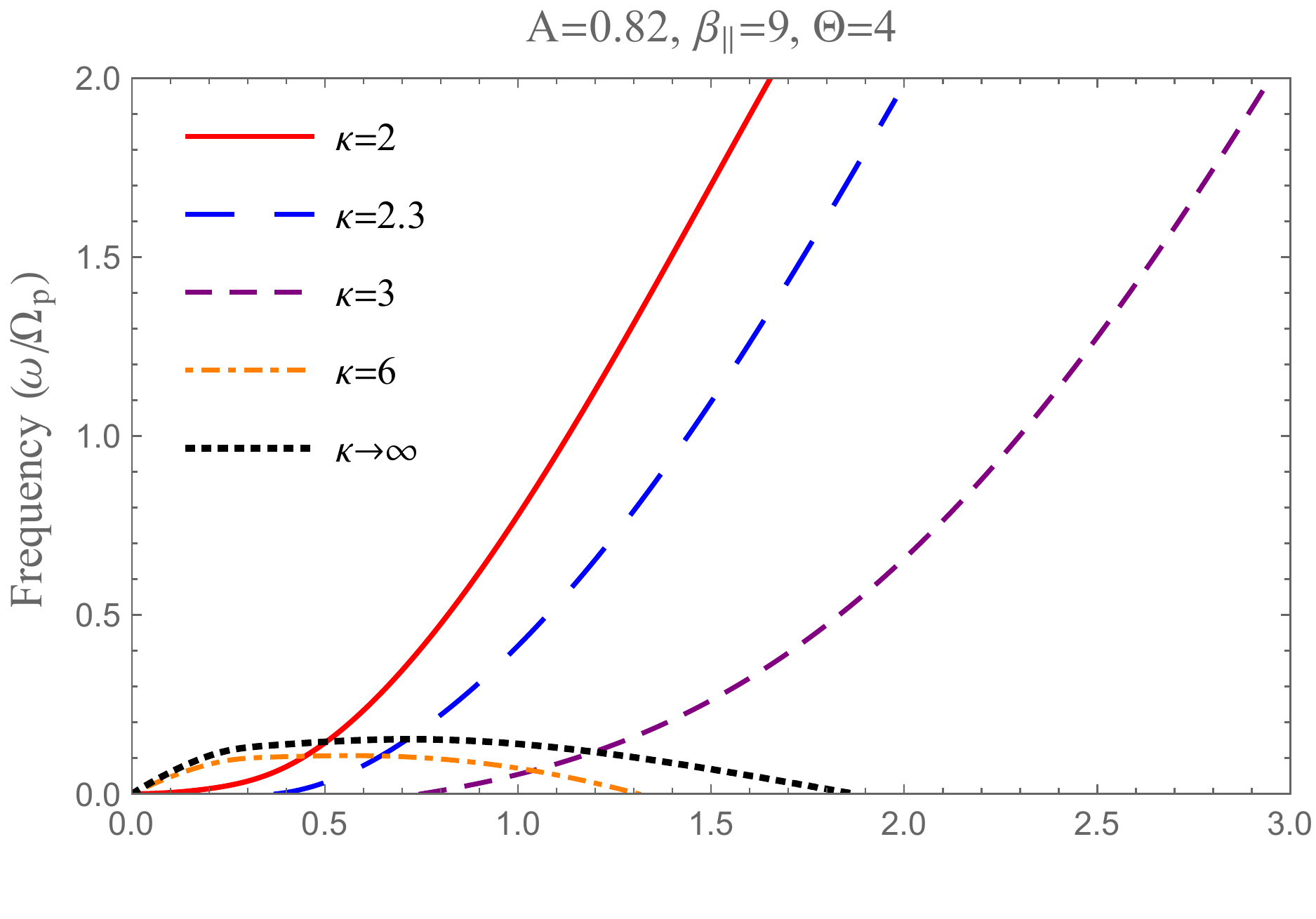}
    \includegraphics[width=80mm]{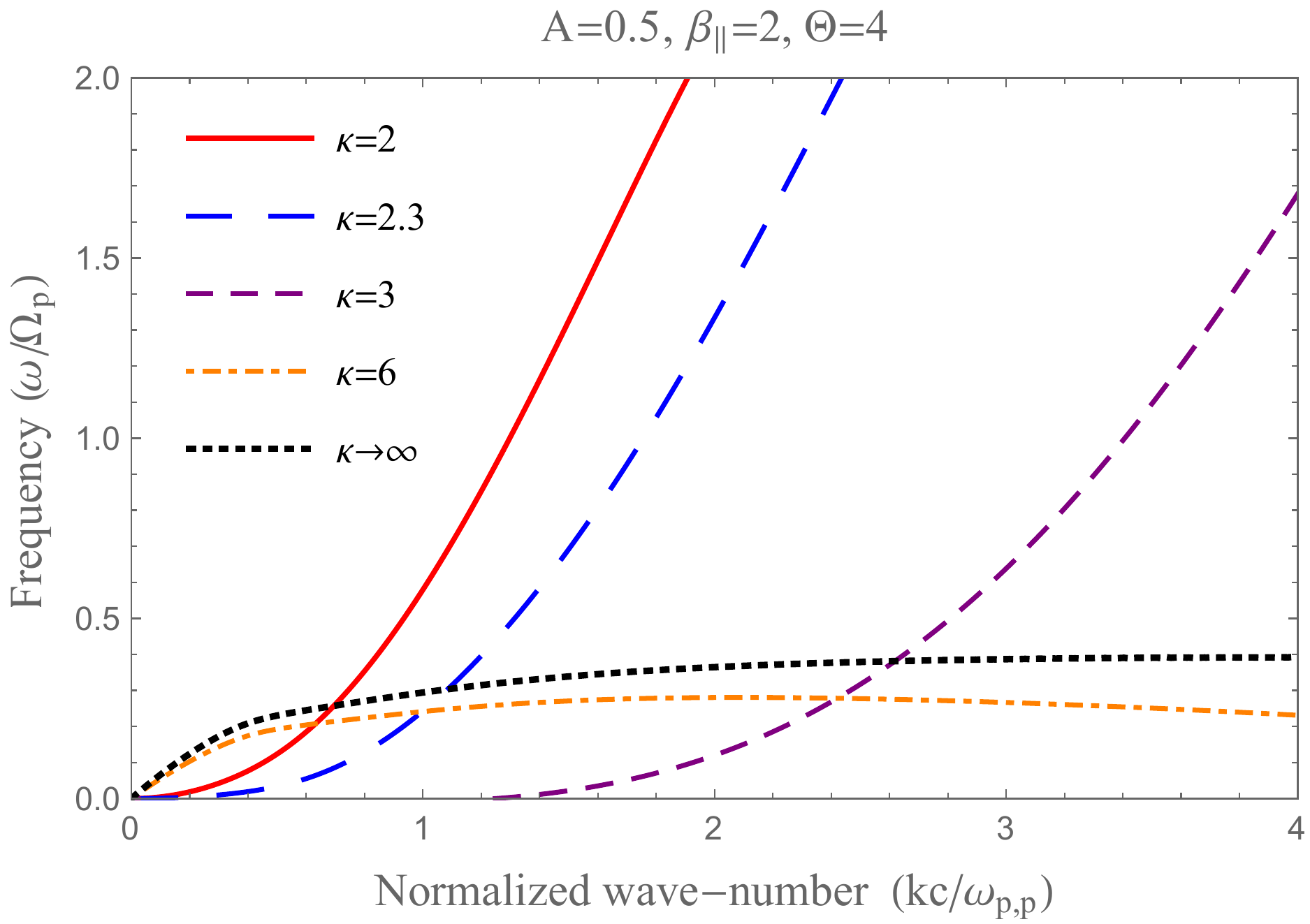}
    \caption{Effects of the suprathermal electrons ($\kappa=$2, 2.3, 3, 6, $\infty$) on the wave-frequency of the 
    EFH instability for the same cases considered in Fig.~\ref{f1}.} \label{f2}%
\end{figure}

\begin{figure}
\centering
    \includegraphics[width=85mm]{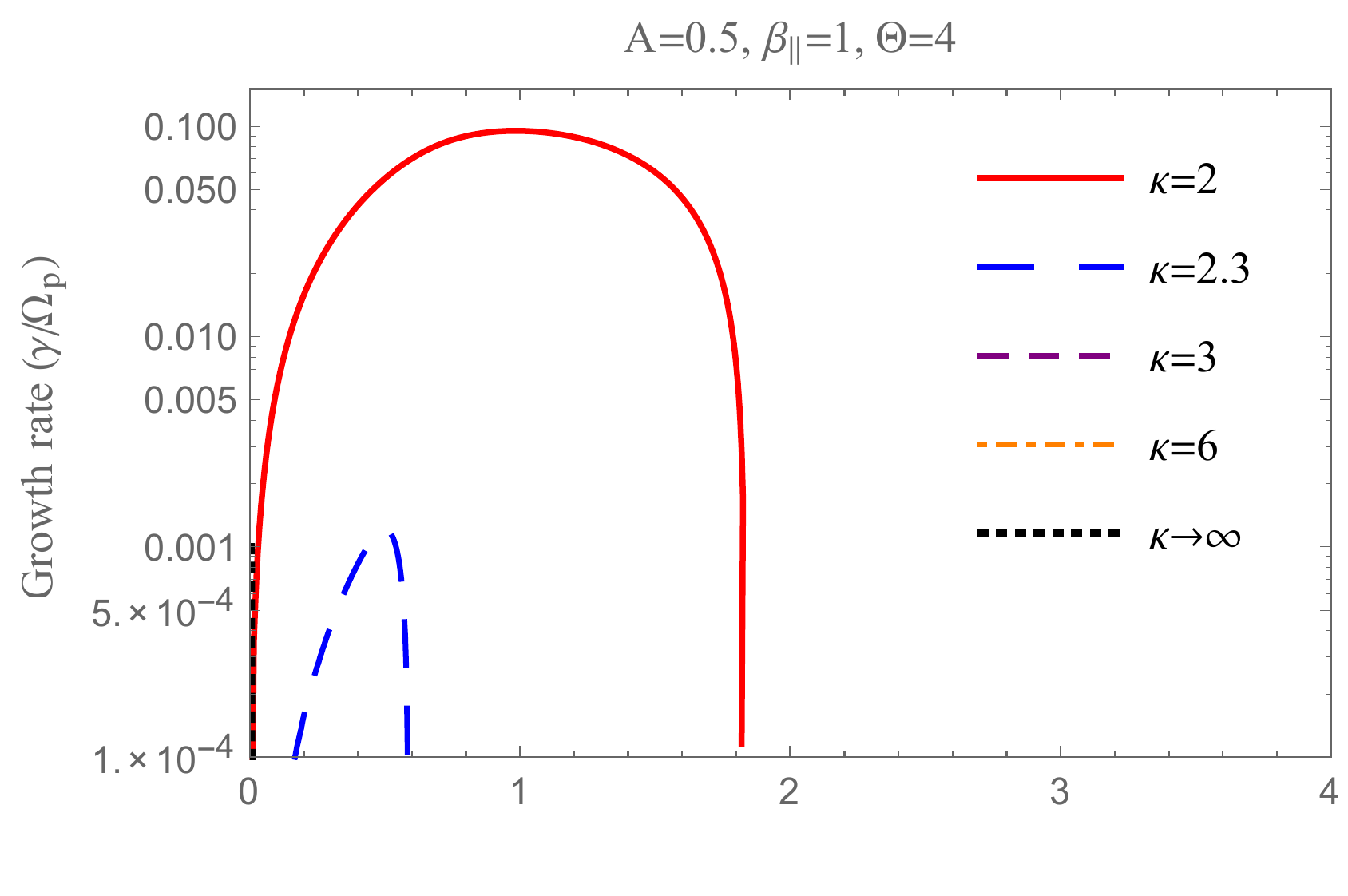}
    \includegraphics[width=85mm]{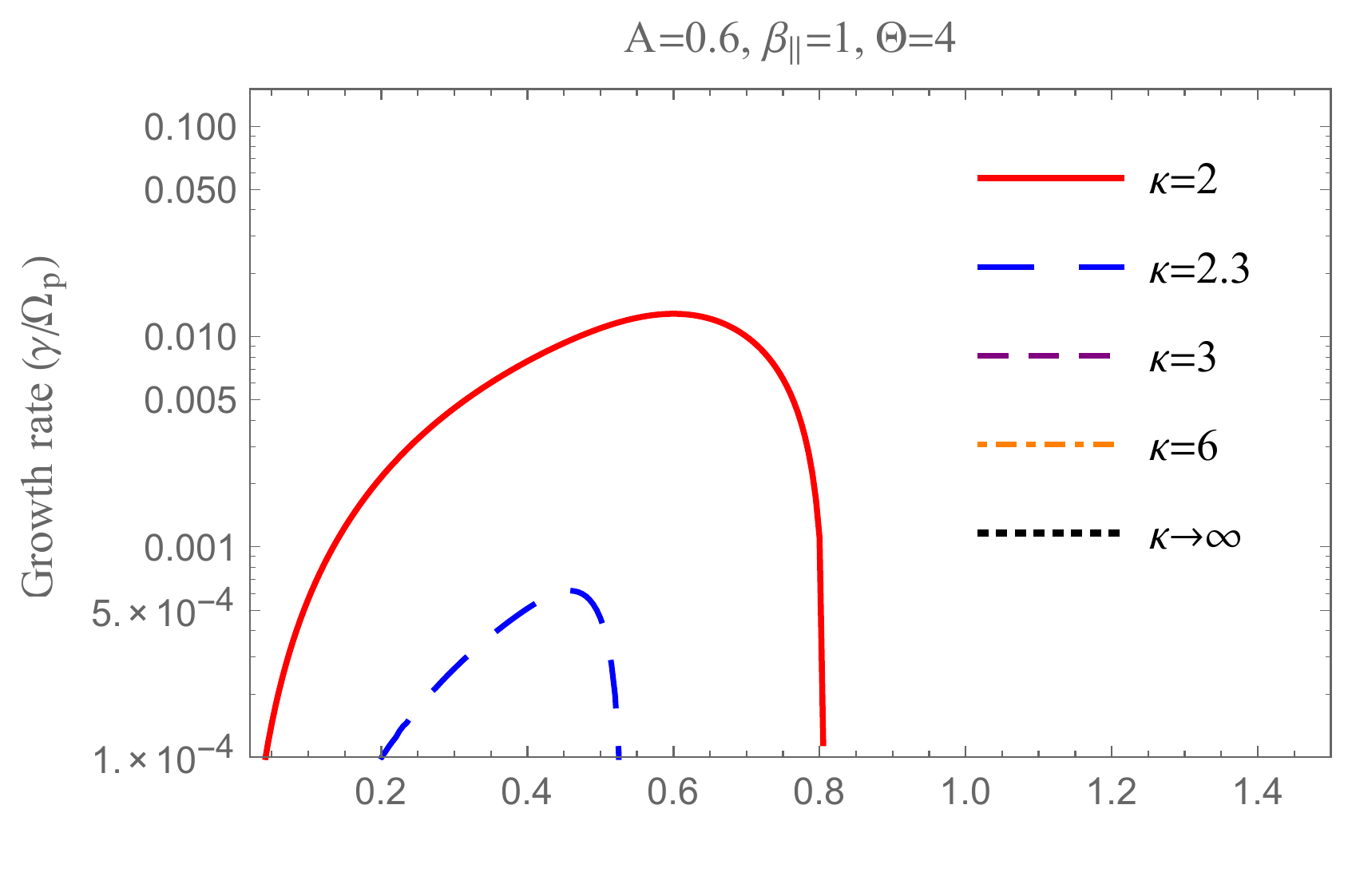}
    \includegraphics[width=85mm]{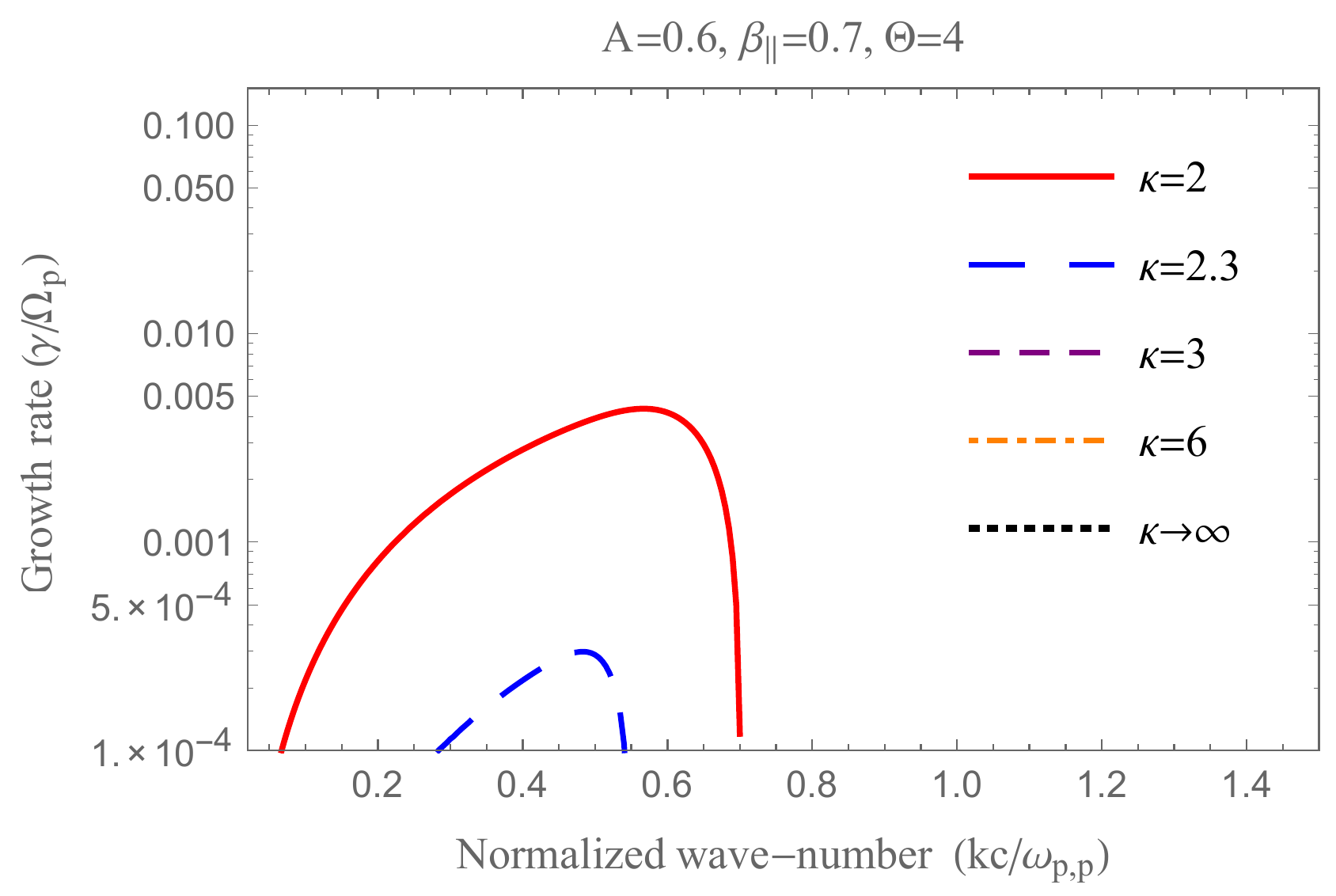}
    \caption{Effects of the suprathermal electrons quantified by the power-index $\kappa=$2, 2.3, 3, 6, 
    $\infty$, on the growth rates of the EFH instability for a lower $\beta_\parallel = 0.6$.} \label{f3}%
\end{figure}

\begin{figure}
\centering
    \includegraphics[width=80mm]{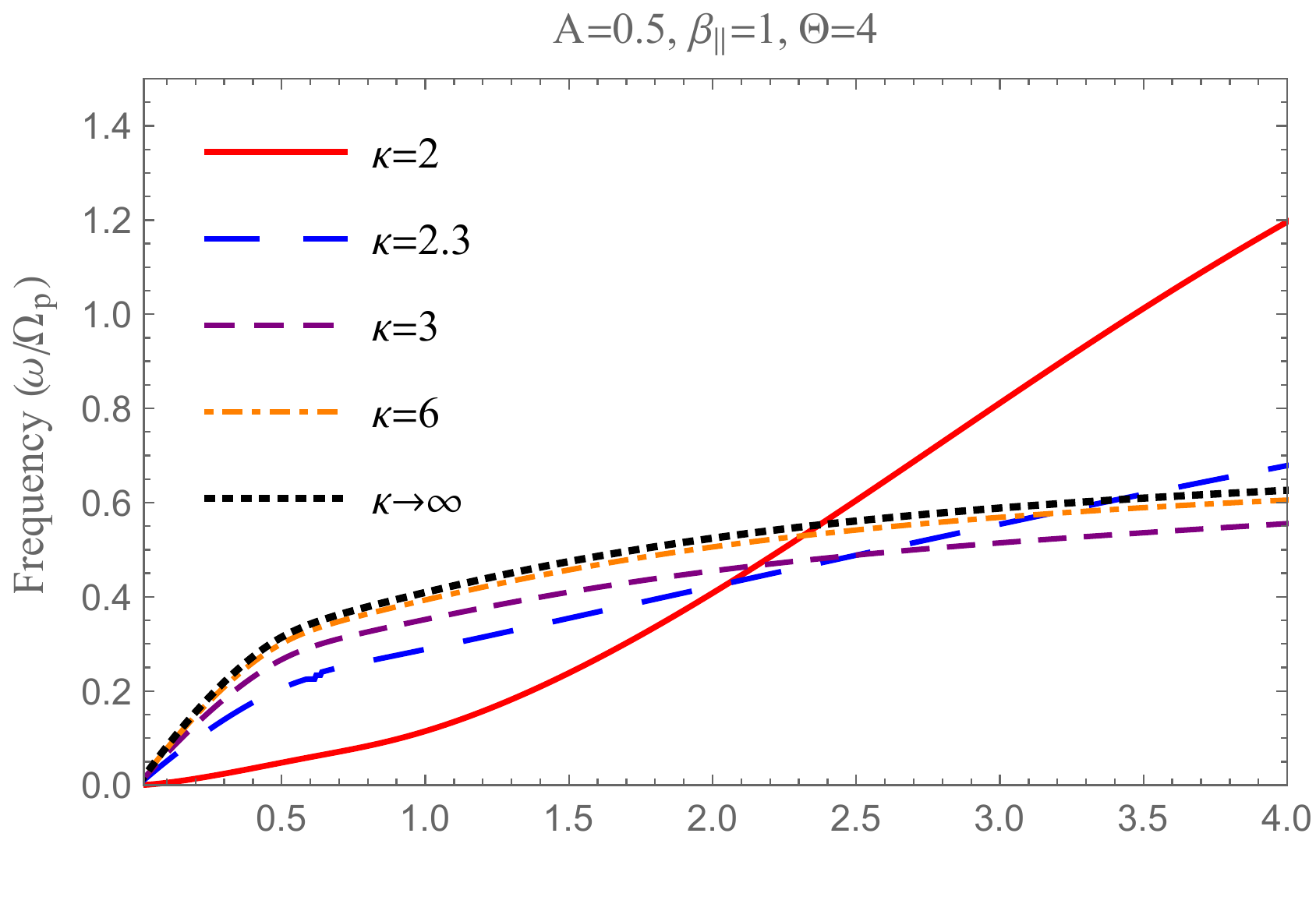}
    \includegraphics[width=80mm]{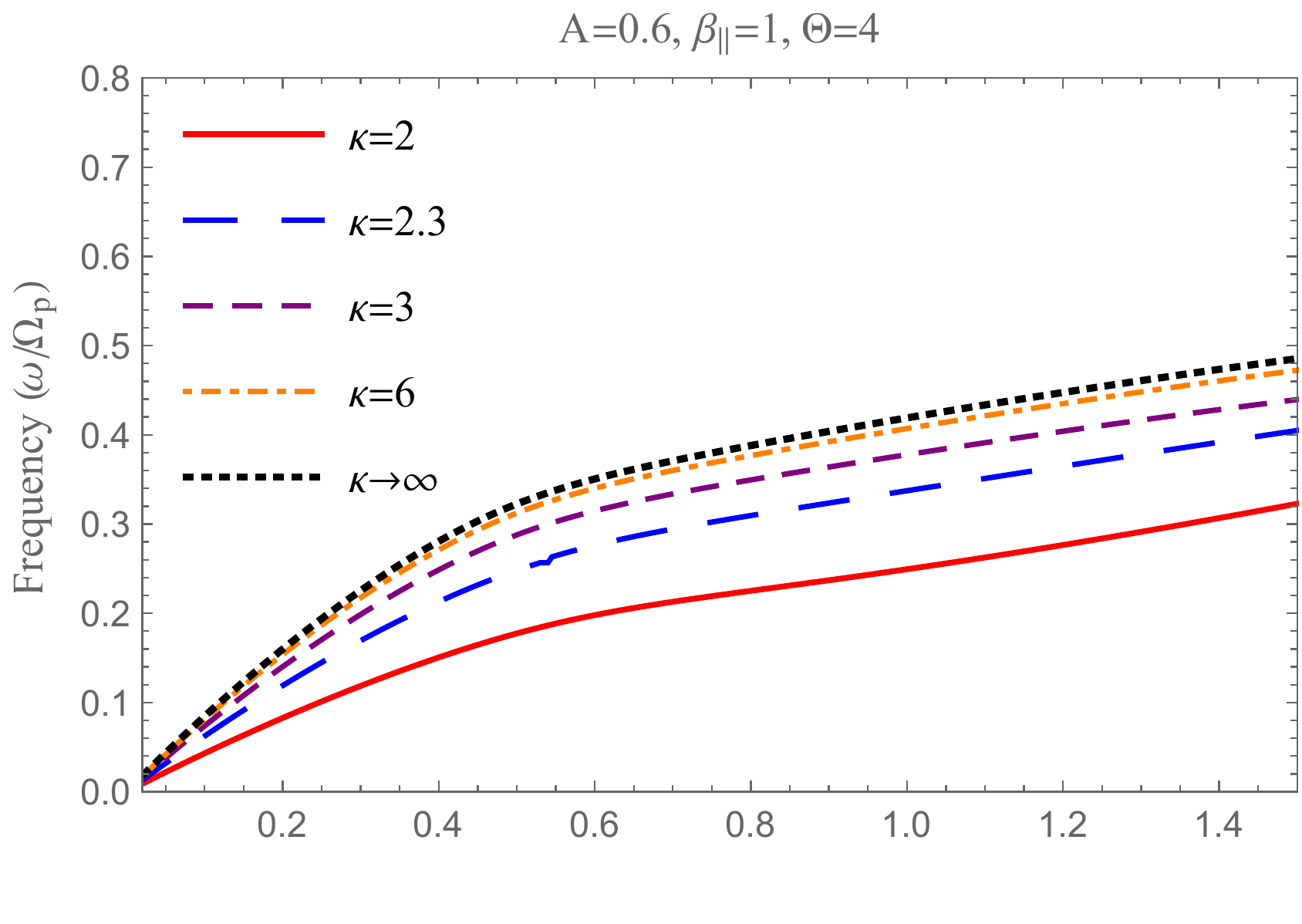}
    \includegraphics[width=80mm]{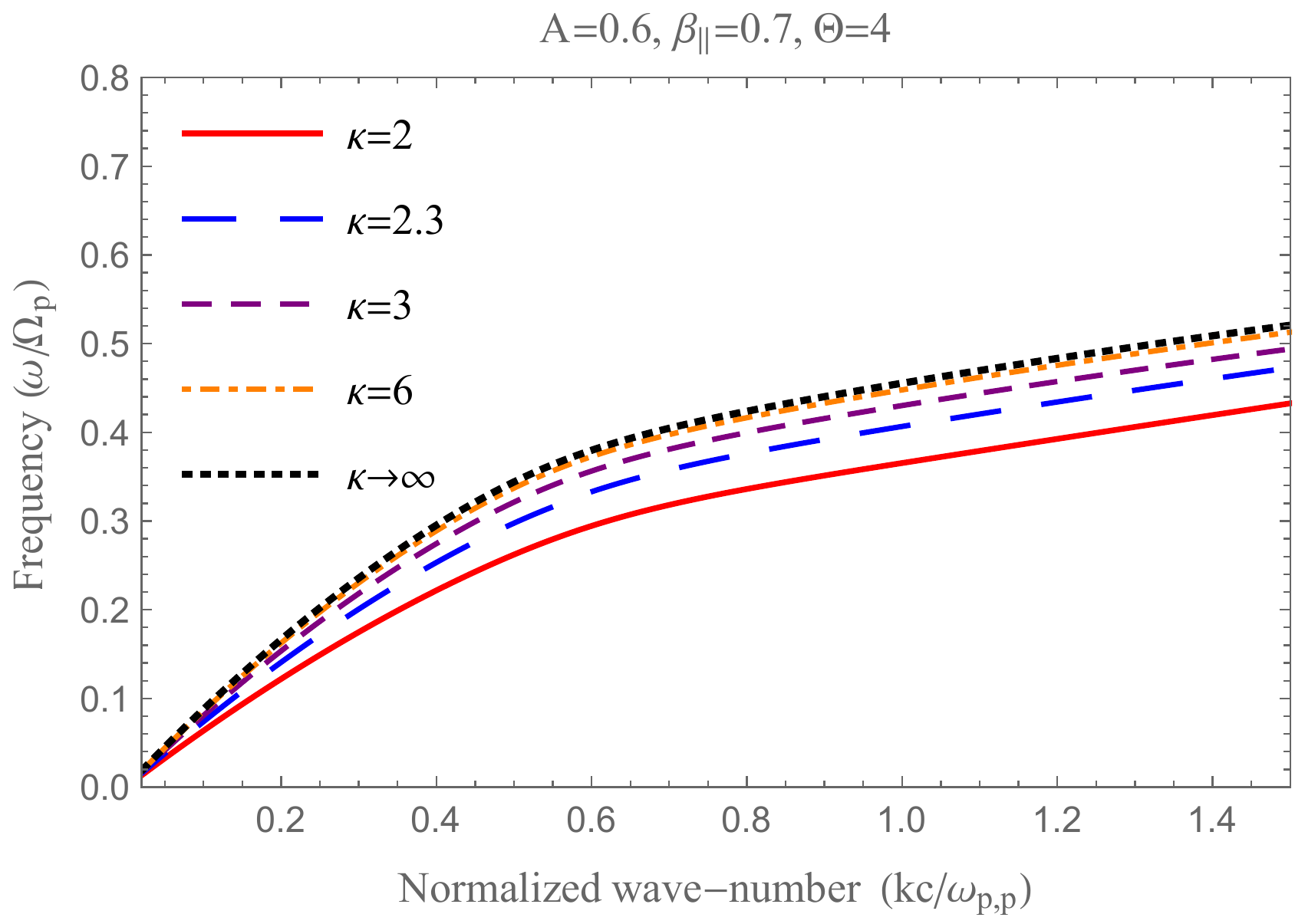}
    \caption{Effects of the suprathermal electrons ($\kappa=$2, 2.3, 3, 6, $\infty$) on the wave-frequency of the 
    EFH instability for the same cases considered in Fig.~\ref{f3}.} \label{f4}%
\end{figure}

Firstly, we examine the growth rates and the wave-frequency of the EFH instability for different plasma regimes conditioned in principal by the (parallel) plasma beta parameter, $\beta_\parallel$,
the electron anisotropy $A$, and the power-index $\kappa$. The regimes identified in Figures~\ref{f1} and 
\ref{f2} are specific to the firehose instability, when a magnetized plasma becomes penetrable 
by the LH electromagnetic fluctuations propagating parallel to the magnetic field with frequencies 
higher than the proton cyclotron frequency. All the unstable modes, i.e., with $\gamma > 0$ in 
Figure~\ref{f1}, exhibit this property that becomes evident in Figure~\ref{f2}, where their 
wave-number dispersion extends to high frequencies exceeding $\Omega_p$.  In the presence of 
suprathermals, i.e., at low values of $\kappa$, the range of unstable wave-nunmbers is 
restrained, but the wave-frequencies and the instability growth-rates are enhanced. These 
effects are in general stimulated by 
increasing the plasma beta parameter $\beta$, the temperature anisotropy and the electron-proton 
temperature contrast $\Theta$. Plots evidencing the influence of $\Theta$ are not shown here, 
but details about this influence are explicitly given in the text. The unstable solutions
displayed in Figures~\ref{f1}-\ref{f4} are obtained for the same value of this parameter, namely, 
for $\Theta=4$ in accordance to the observations in the slow solar wind \citep{Ne1998}.
At higher values of $\kappa$
the instability conditions may be not satisfied and the electromagnetic modes are damped,
e.g., $\gamma < 0$ for $\kappa \geqslant 6$ in Figure~\ref{f1}, middle and bottom panels.
For these modes, the wave-frequency dispersion curves displayed in Figure~\ref{f2} have 
a different allure, showing an asymptotic increase similar to the ion (proton)
cyclotron modes with frequencies always smaller than  $\Omega_p$. These are LH modes damped 
by the protons and therefore limited only to the large (proton) scales. 
At smaller scales controlled by the electrons (higher wave-numbers) these 
modes change (mode conversion) to RH polarization (i.e., the wave-frequency displayed in 
Figure~\ref{f2} becomes negative) which is more specific to the electron whistlers.   

In Figures~\ref{f3} and \ref{f4} we show that these LH-polarized modes with a wave-number 
dispersion resembling that of the electromagnetic ion cyclotron (EMIC) modes can be destabilized
by the anisotropic bi-Kappa distributed electrons, see middle and bottom panels. 
This is a new regime of the EFHI 
destabilizing only the low-frequency branch of the LH modes with wave-frequency showing 
an asymptotic increase of their wave-frequencies but remaining always below  $\Omega_p$. 
To establish this regime the kinetic effects of the electrons are also tempered by considering lower 
values of plasma beta, and the instability is triggered only by the anisotropic 
distributions with sufficiently low $\kappa$, e.g., $\kappa < 3$ in Figures~\ref{f3} and \ref{f4}. 
Furthermore, in this case, both the (maximum) growth-rates and the range of the unstable 
wave-numbers are considerably enhanced by increasing the presence of suprathermals, i.e.,
lowering the values of $\kappa$. Again, these features seems to be more specific to the instability 
of the cyclotron modes \cite{Shaaban2016JGR}. 
The transition between the classical EFH solutions (exemplified in Figures~\ref{f1}
and \ref{f2}) and the new regime of a low-frequency EFHI is suggestively shown by the 
top panels in Figures~\ref{f3} and \ref{f4}.
In these panels we have unstable solutions specific to both these regimes: the solid-line 
solution obtained for $\kappa = 2$ is a classical firehose, while the next long-dashed-line 
solution obtained for $\kappa = 2.3$ is already more specific to the new regime of EFHI.   
In this case it is only the power-index $\kappa$ that may switch between these two regimes,
but a direct comparison of the other plasma parameters in Figures~\ref{f1}-\ref{f4}, clearly
shows that these regimes are also conditioned by the temperature anisotropy, the plasma beta,
and the temperature contrast between electrons and protons.



\begin{figure}
\centering
    \includegraphics[width=80mm]{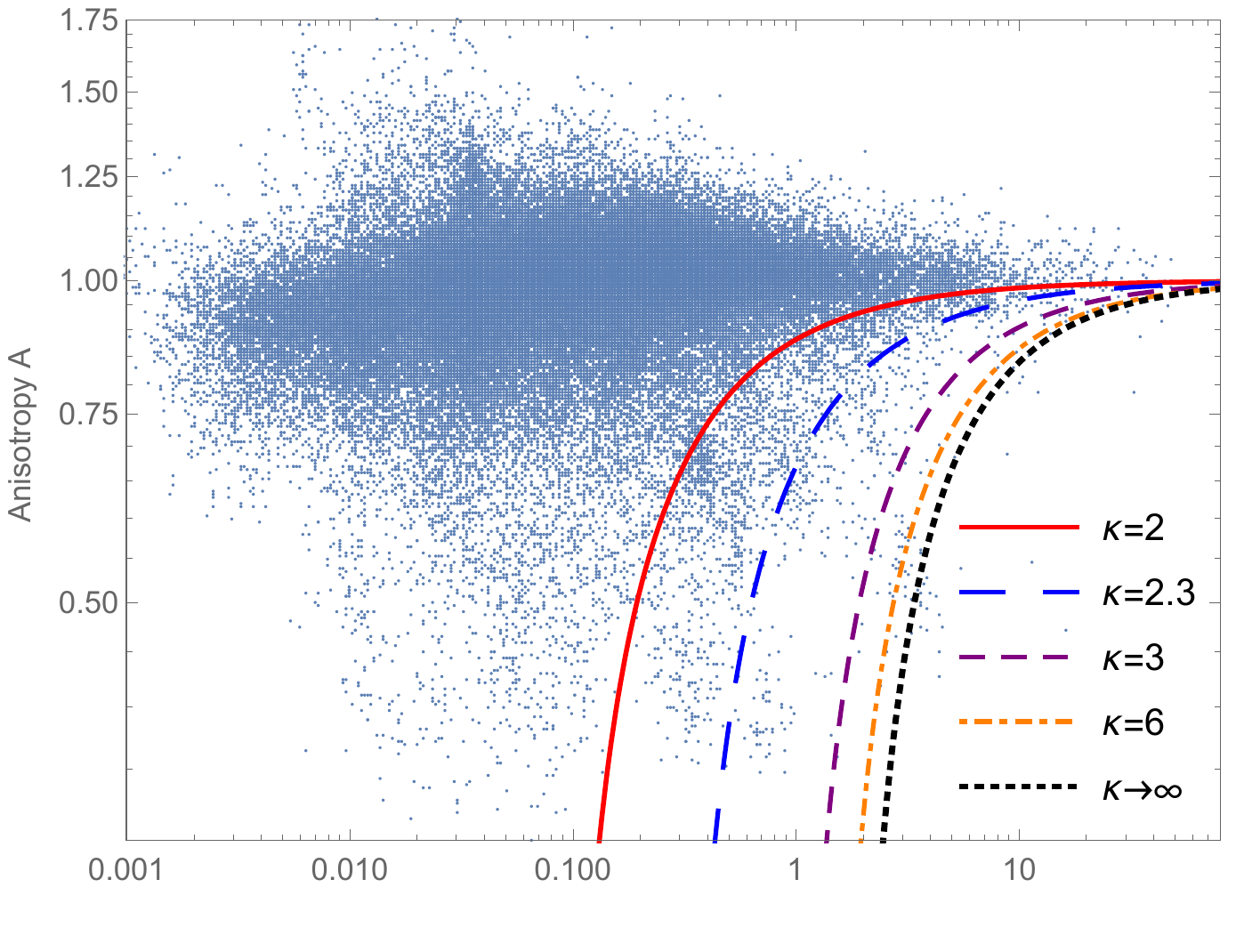} \\
    \includegraphics[width=80mm]{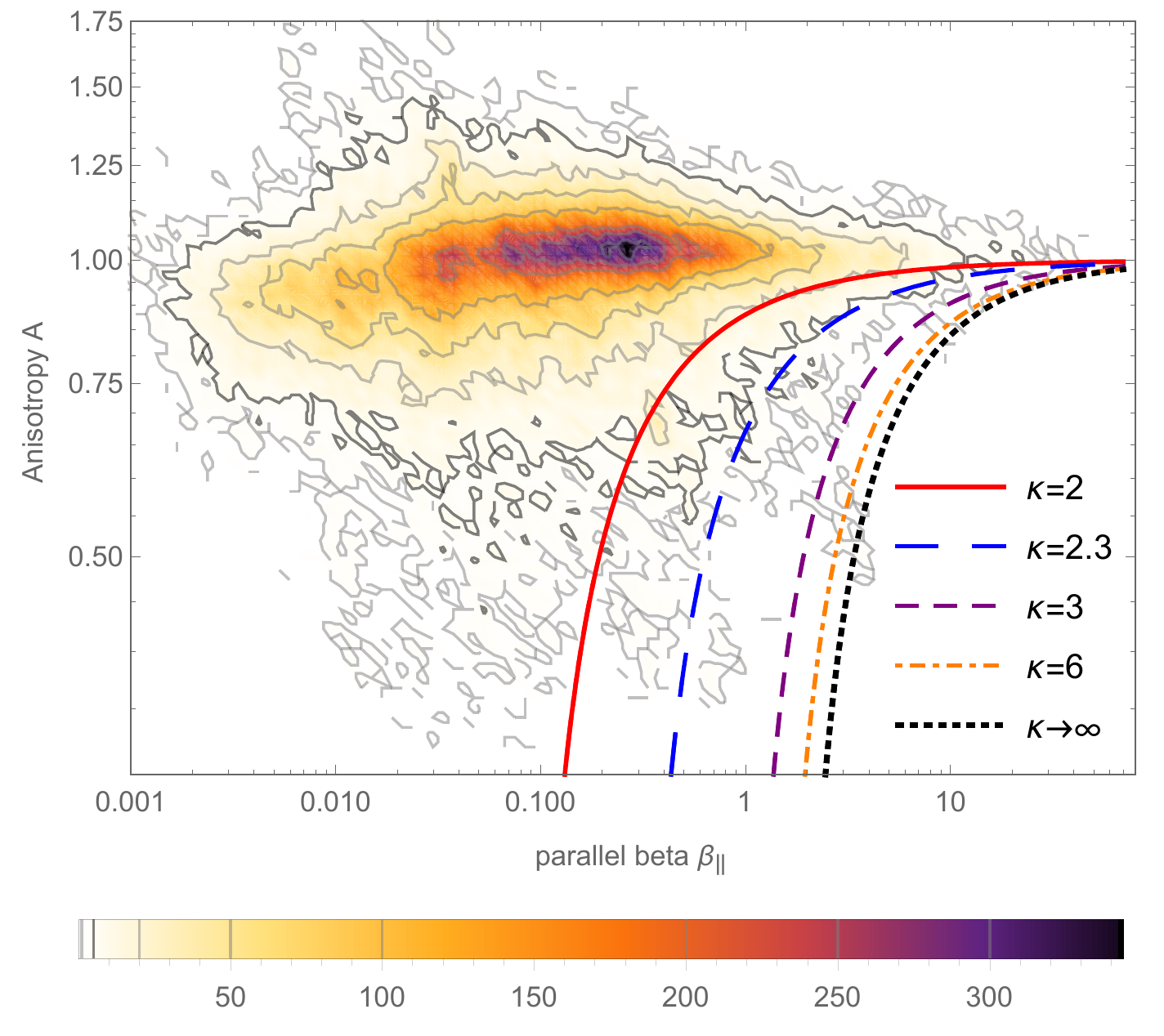}
    \caption{Comparison of the anisotropy thresholds (\ref{e14}) for maximum growth rates $\gamma_m / \Omega_p 
		= 10^{-3}$ with the temperature anisotropy measured in the solar wind which is displayed using a scatter 
		plot data in the top panel and a histogram data in the bottom panel.} \label{f5}%
\end{figure}

In the second part of this section we analyze the anisotropy thresholds of the instability.
These thresholds represent plasma conditions associated with given values of the maximum growth-rate, 
usually small values, e.g., $\gamma_m /\Omega_p = 10^{-2}$, $10^{-3}$, approaching the marginal 
stability $\gamma_m/\Omega_p \to 0$. In Figure~\ref{f5} we display the instability thresholds 
associated with $\gamma_m/\Omega_p=10^{-3}$ and derived for different values of the electron 
power-index $\kappa$. These are isocontours of the electron temperature anisotropy $A$ as a 
function of the parallel electron plasma beta $\beta_{\parallel}$, fitted to an inverse
correlation law of the form \citep{Ga1994,Ga1998}
\begin{equation}
A=1+\frac{a}{\beta _\parallel^b}. \label{e14}
\end{equation}
The values obtained for the fitting parameters $a$ and $b$ can be found in Table~\ref{t1}.
For the plasma beta parameter we consider an extended range of values $0.1<\beta_{\parallel}<50$ 
relevant for the electron halo populations in the solar wind \citep{St2008}. Higher 
values of $\beta_{\parallel}$, associated with hotter plasmas or less intense magnetic 
fields, means lower deviations from isotropy to trigger the instability. 
The effects of suprathermal electrons is reconfirmed here by a systematic stimulation of 
the (maximum) growth-rates with decreasing $\kappa$. As a consequence, 
the anisotropy thresholds are found to be markedly lowered in the presence of suprathermals,
and this effect may be enhanced by increasing the temperature contrast between electrons
and protons. Larger variations of the anisotropy thresholds are obtained at lower values
of $\kappa$.

The instability thresholds are compared in Figure~\ref{f5} with the observational data of 
the electron halo populations in the slow solar wind ($v< 500$ km/s), which are displayed 
in the top panel as a scatter plot, and in the bottom panel as a histogram counting the 
number of events within a color logarithmic scale. This data set comprises more than 120 000 
events detected by three space missions (Helios 1, Cluster II, and Ulysses) at different 
heliocentric distances (in the interval 0.3--3.95~AU) in the ecliptic. The details about 
the electron analyzers used by these missions, and the methods of correction and 
reconstruction of the 3D velocity distribution functions can be found in \cite{St2008}.
These authors have used the same set of events to analyze the temperature anisotropy
of the main electron populations, namely, the thermal core and suprathermal halo, and the 
most plausible constraints exercised on their temperature anisotropy by different physical 
mechanisms, e.g., collisions and kinetic instabilities. However, \cite{St2008} have limited
to investigate in detail only the bi-Maxwellian core anisotropy finding that the particle-particle 
collisions still may have an effect to constrain low levels of anisotropy, while the kinetic
instabilities occur for larger deviations from isotropy, which exceed their thresholds. 
Indeed, the instability thresholds predicted by a bi-Maxwellian model are found to shape very 
well the limits of the core anisotropy, but the same thresholds cannot explain the limits observed 
for the temperature anisotropy of suprathermal electrons. Figure~6 from \cite{St2008} presents
such a comparison between the observational data and the EFHI thresholds predicted by a bi-Maxwellian
approach, which is also reproduced here in Figure~\ref{f5} by the dotted line corresponding to $\kappa \to \infty$.
Moreover, in Figure~\ref{f5} we show that this disagreement may be resolved by the instability thresholds
derived for bi-Kappa models which are more appropriate to describe the suprathermal electrons. 
The instability thresholds are markedly changed with decreasing the power index $\kappa$ and 
for lower values of $\kappa$ these thresholds are approaching the limits of the temperature 
anisotropy observed in the solar wind. What we found even more interesting is that the instability 
thresholds also shape very well the isocontours of the observational data, counting the number 
of events in the bottom panel. 
Developing for large temperature anisotropies exceeding these thresholds, the FHI dissipates
the free energy and enhances the electromagnetic fluctuations, which may also prevent the 
anisotropy to grow by scattering particles back towards quasi-equilibrium states. 
Given that suprathermal electrons in the solar wind are practically collisionless, such a 
good agreement between the instability thresholds predicted by the kinetic theory and the limits of 
the temperature anisotropy reported by the observations represents an important confirmation
on the role played by the FHI instability in the relaxation process. 

\begin{table}
\centering \caption{Fitting parameters for thresholds $\gamma_{\rm m}/\Omega_p = 10^{-3}$} \label{t1}
\begin{tabular}{c c c c c c c c }
\hline
 Fit & $\kappa=2$& $\kappa=2.3$ & $\kappa=3$ & $\kappa=6$  &$\kappa\rightarrow\infty$ \\
\hline
$a$        &-0.1196 & -0.3304 & -0.9615 & -1.3733 & -1.7950 \\          
$b$        & 0.8708 & 0.8996 & 1.0009 & 1.0030 & 1.0456  \\
\hline
\end{tabular}
\end{table}


\section{Discussions and conclusions}

In this paper we have proposed a refined theory of the electron firehose instability 
in anisotropic Kappa distributed plasmas, which provide a new and, in our opinion, valuable evidence 
of an extended implication of this instability in the relaxation of the temperature anisotropy 
in collision-poor plasmas from space. 
Our present study is particularly motivated by the solar wind observations which do not 
confirm the indefinite increase of temperature predicted by the solar wind expansion in the 
direction parallel to the interplanetary magnetic field, but reveal very clear bounds for the 
temperature anisotropy of plasma particles. Previous studies have focused to the
thermal (core) populations of electrons and protons, using standard bi-Maxwellian approaches, 
and have shown that large deviations from isotropy are constrained by the kinetic instabilities
\citep{Hellinger06, St2008}. However, the same bi-Maxwellian is not appropriate to describe 
suprathermal populations and their anisotropy, and cannot prescribe accurately the resulting 
instabilities and their back reaction on these populations.

Here we have assumed the anisotropic electrons well reproduced by the bi-Kappa distribution 
function, which is the empirical model invoked by \cite{St2008} to describe the velocity distribution 
of suprathermal electrons in the solar wind. In addition, the temperature of the suprathermal 
population is considered dependent on the power-index $\kappa$, enabling us a realistic 
interpretation of the suprathermals and their effects (theoretical and observational arguments 
are detailed in the Introduction). The results of our present study contrast with 
those provided by \cite{La2009} and \cite{La2011}, who studied the same EFHI but driven by 
bi-Kappa electrons with a $\kappa$-independent temperature. These differences are made clear in 
the discussion from this section. 

Two distinct regimes of the EFHI are identified in Section~3, and these regimes are differentiated 
by the wave-number dispersion laws (curves) obtained for the wave-frequency and growth-rate
of the instability. Thus, more specific to the EFHI are the unstable LH-polarized modes 
exemplified in Figures~\ref{f1} and \ref{f2} with frequencies that can significantly exceed 
the proton cyclotron frequency $\Omega_p$. For this regime to be established, the kinetic 
free energy of the anisotropic electrons
must be sufficiently large, and this usually means a high enough plasma beta or a large
anisotropy. If damped, these modes cannot extend above $\Omega_p$ and 
their wave-number dispersion keeps the aspect of low-frequency EMIC modes in the absence of 
kinetic anisotropies. At higher wave-numbers (lower scales) these damped modes 
can change their polarity converting to the branch of RH-polarized modes (whistlers). 
These electromagnetic modes with a wave-number dispersion resembling that of the EMIC modes,
i.e., with wave-frequency increasing asymtotically to $\Omega_p$, can be driven unstable 
by the EFHI for conditions approaching marginal stability. A few cases relevant for this new regime are presented in Figures~\ref{f3}
and \ref{f4}, with mention that top panels include unstable solutions representative for a
transition between these two distinct branches of the EFHI. 

We should observe that considering plasma parameters with values typical 
for the solar wind conditions, e.g., in Figures~\ref{f1}-\ref{f4}, the EFHI develops only in
the presence of suprathermal electrons, i.e., for finite values of $\kappa$, 
while for (bi-)Maxwellian limit $\kappa \to \infty$ these modes are damped. 
Increasing the presence of suprathermal populations (by lowering $\kappa$) 
has opposite effects on the wave-frequency of the unstable modes, which become evident if we compare 
for instance Figures~\ref{f2} and \ref{f4}. However, the EFHI is clearly stimulated by the 
suprathermal electrons, which enhance the (maximum) growth-rates in both these two regimes. 
Noticeable is the significant increase shown by the growth-rates for conditions approaching 
the marginal stability (Figure~\ref{f3}), which can also explain the significant decrease of 
the instability thresholds shown in Figure~\ref{f5}. These thresholds are markedly lowered with 
decreasing $\kappa$, and for lower values of $\kappa$ they shape very well the limits 
of temperature anisotropy
reported by the observations in the solar wind. In the previous studies involving bi-Kappa 
electrons with a $\kappa$-independent temperature, e.g., in \cite{La2009, La2011}, the existence 
of these two distinct regimes was not mentioned, and, in contrast to our present results, 
the suprathermals were found inhibiting the EFHI, and departing the instability thresholds 
from the limits of temperature anisotropy in the solar wind.


Our present results strongly suggest that the EFHI may efficiently constrain the 
temperature anisotropy of the suprathermal electrons in the slow wind, complementing the results by 
\cite{St2008}, which showed the same effects of this instability on the core electrons.
A good agreement between the instability thresholds and the bounds of the temperature
anisotropy measured in the solar wind is conditioned by a proper modelling of the velocity
distributions in accord to the observations. In conclusion, the EFHI can be considered a 
plausible mechanism of electron energy transfer between the directions parallel and 
perpendicular to the uniform magnetic field. From an extended perspective, we can further claim 
that the resulting low-frequency fluctuations can establish an energy transfer from small 
to large scales, namely from the electrons, especially the energetic or suprathermal electrons 
which carry the main heat flux in the solar wind, to the resonant protons. Although suprathermal
populations are not easily captured in numerical experiments, it becomes however clear that 
our present results provide valuable premises that may stimulate new and advanced simulations 
to confirm these mechanisms.

\section*{Acknowledgments}
The authors acknowledge support from the Katholieke Universiteit Leuven,
Ruhr-University Bochum, and Alexander von Humboldt Foundation. 
These results were obtained in the framework of the projects
GOA/2015-014 (KU Leuven), G0A2316N (FWO-Vlaanderen), and C 90347
(ESA Prodex 9). The research leading to these results has also
received funding from the European Commission's Seventh Framework 
Programme FP7-PEOPLE- 2010-IRSES-269299 project-SOLSPANET
(www.solspanet.eu). S.M. Shaaban would like to thank the Egyptian Ministry 
of Higher Education for supporting his research activities.

\label{lastpage}
\end{document}